\documentclass[prb,aps,floats,amssymb,showkeys,showpacs,preprint,
superscriptaddress]{revtex4}
\usepackage{graphicx} 
\usepackage{graphics}
\usepackage{amsmath}
\usepackage{epsfig,bm}
\usepackage{rotating}
\begin{document}

\title{Sums of permanental minors using Grassmann algebra}

\author{P. Butera}
\email{paolo.butera@mib.infn.it}
\affiliation
{Dipartimento di Fisica Universita' di Milano-Bicocca\\
and\\
Istituto Nazionale di Fisica Nucleare \\
Sezione di Milano-Bicocca\\
 3 Piazza della Scienza, 20126 Milano, Italy}

\author{M. Pernici} 
\email{mario.pernici@mi.infn.it}
\affiliation
{Istituto Nazionale di Fisica Nucleare \\
Sezione di Milano\\
 16 Via Celoria, 20133 Milano, Italy}

\date{\today}
\begin{abstract}

  We show that a formalism proposed by Creutz to evaluate Grassmann integrals
provides an algorithm of complexity $O(2^n n^3)$ to compute
  the generating function for the sum of the permanental minors of a matrix
  of order $n$.
  This algorithm improves over the Brualdi-Ryser formula, whose
  complexity is at least $O(2^{\frac{5 n}{2}})$.
  In the case of a banded matrix with band width $w$ and 
  rank $n$ the complexity is $O(2^{min(2w, n)} (w + 1) n^2)$.

  Related algorithms for the matching and independence polynomials
  of graphs are presented.

\end{abstract}
\maketitle

\section{ Introduction} 
Let $G$ be an undirected graph with $E$ edges and $v = |V|$ vertices.
A matching of $G$ is a set of pairwise disjoint edges.
The  matching generating polynomial of  $G$,  
i.e.  the generating function
 of the number $N(i)$ of different  matchings of  $G$ 
containing $i$ edges, defined by
\begin{equation}
M(t) = \sum_{i=0}^{[v/2]} N(i)t^i
\end{equation}
first appeared in
combinatorics as the ``rook'' polynomial\cite{riordan}. It was then
introduced in statistical physics\cite{HL70, HL72} for 
the study of the monomer-dimer system on a lattice,  in theoretical
chemistry\cite{hosoya71} to compute the ``Hosoya index''
$Z(G)= M(1)$,
i.e. the total number of matchings of $G$.

$M(-1)$ on a lattice graph can be interpreted as the Witten index of a
supersymmetric dimer model defined on that lattice\cite{fse}.

For a generic graph, the best current algorithms for computing the
matching generating polynomial are based\cite{godsil} on recurrence relations.
Alternatively, for a bipartite graph the coefficients of the matching
polynomial can be computed as the sum of the permanental minors of the
 reduced adjacency matrix of the graph (defined as the submatrix of the
adjacency matrix from the even to the odd vertices).  The best current
algorithm for computing the sum of the permanental minors is the
Brualdi-Ryser formula\cite{Brualdi}.
Notice that even in the case of the permanent of banded matrices the
complexity of currently used algorithm is in general exponential.
In Ref.[\onlinecite{tw}] it was shown that for banded matrices which are block 
factorizable the permanent can be computed in polynomial time.

 The study of graph matchings can be naturally\cite{fisgreen,samuel} formalized
by introducing anticommuting variables, so that the edges in a
matching cannot overlap due to the Pauli exclusion principle.  
Creutz introduced an efficient algorithm for Grassmann integration\cite{creutz}.
In Ref.[\onlinecite{FJ}] th Creutz algorithm is applied
to a graph coloring problem.

In this paper, we shall present a simplified form of Creutz algoritm,
in which Grassmann integration reduces to simple polynomial
manipulations.

A hard object is represented as a product of even
elements $\eta_i$ of a Grassmann algebra, associated to the nodes $i$
of the graph on which the objects lie.  The $\eta_i$ elements are
commuting and nilpotent and are represented as products 
$\eta_i = \bar{\theta}_i\theta_i $ of anticommuting variables 
$\theta_i,\bar{\theta}_i$ .
  In the
case of dimer systems, this notation was introduced\cite{fisgreen,hayn} as a
starting point to deduce the free-fermion interpretation of the
close-packed dimer model on planar lattices.

The generating function that counts the hard objects is a Grassmann
integral of a product on these objects. 

Our algorithm to compute the generating function of the sums of the
permanental minors of a matrix of order $n$ has time complexity $O(2^n
n^3)$, while for the Brualdi-Ryser formula the complexity is larger
than $O(2^{\frac{5 n}{2}})$.  In the case of banded matrices with a
fixed band-width $w$ the former algorithm has quadratic complexity in
$n$, $O(2^{2w} (w + 1) n^2)$, the latter exponential.

In the case of dimers, the $\eta$ elements are associated to the
end-point of the dimer and we obtain efficient prescriptions to
compute the matching polynomial for both bipartite and non-bipartite
graphs.

Another important graph polynomial associated to $G$, the independence
polynomial, $I(t)=\sum_{i=0} a(i) t^i$, with $a(i)$ the number of
independent subsets of $i$ vertices in $V$, can be similarly
derived. In this case the hard object is represented by $\eta$
elements associated to the edges adjacent to a vertex.

Appendix A tabulates the values of the Witten indices in the cases of
square and hexagonal lattices with relatively large size, for some pf
which we disagree with the results of the calculation in
Ref.[\onlinecite{Eerten}].

We provide an implementation of these algorithms in Python;
examples of its usage are given in Appendix B.

\section{An algebraic formalism for counting hard objects on a graph}

Define a ``hard object'' $a$ on a graph $G$ with set of vertices $V$
and of edges $E$ as a subset $V_a$ of the vertices in $V$. This is a
generalization of the notion of dimer.  Configurations of two or more
hard objects onto $G$ are admissible provided they have no common
vertices (i.e. the vertex subsets of the various objects are
``independent'').

Let us associate to each object $a$ the expression 
\begin{equation}
O_a = 1 + w_a\prod_{i \in V_a} \eta_i
\label{obj}
\end{equation}
\noindent
where $w_a$ is a weight factor, and the product runs on a set of elements
 $\eta_i=\bar \theta_i\theta_i$, where $\theta_i$, $\bar{\theta}_i$
 are Grassmann anticommuting variables.

Define
\begin{equation}
<A> = \int \prod_{i=1}^v d\theta_i d \bar \theta_i
exp(\sum \bar \theta_i \theta_i) A
\end{equation}
where the Berezin integration\cite{berezin} over anticommuting variables
is used.

The $\eta$-elements satisfy the rules
\begin{equation}
\eta_i^2 = 0
\label{forms1}
\end{equation}
\begin{equation}
\eta_i \eta_j = \eta_j \eta_i
\label{forms2}
\end{equation}
\begin{equation}
<\eta_{i_1}...\eta_{i_k}> = 1   
\label{forms3}
\end{equation}
when $i_1,.., i_k$ are all distinct.
Consider now the product of all admissible objects onto the graph $G$
\begin{equation}
Z_G = <\prod_{a} O_a >
\label{prodobj}
\end{equation}

One can write 
\begin{equation}
Z_G = <\prod_a (1 + w_a\prod_{i \in V_a} \eta_i) >
  = \int d\theta d\bar \theta exp(S)
\end{equation}
with
\begin{equation}
    S = \sum_{i \in V} \bar \theta_i \theta_i 
+ \sum_a w_a \prod_{i \in V_a} \bar{\theta}_i\theta_i
\end{equation}

If $w_a=t$ for all $a$, where $t$ is a variable, $Z_G(t)$ is the
generating function of the number of ways to settle the hard objects onto
the graph.

Let us observe that after performing a partial product $\prod_{b}
O_b$, if an element $\eta_i$ does not occur in the remaining products
in $Z_G$, then one can replace $\eta_i$ with $1$. This reduces the
number of possible monomials, thus simplifying the product.  To save
memory and improve performance, the product should be ordered in such
a way that only few elements $\eta_i$ are present for any partial
product.
This algorithm is a simplified version of Creutz algorithm\cite{creutz};
since the Grassmann variables $\theta_i$ and
$\bar{\theta_i}$ appear only in $\eta_i$, we can avoid introducing
the Fock space for fermionic operators and use only simple
polynomial manipulations.

\subsection{Sums of permanental minors}

The permanent of a $n\times n$ matrix $A$ is the coefficient\cite{Percus} of
the $x_1...x_n$ monomial in

\begin{equation}
\prod_{i=1}^n \sum_{j=1}^n A_{ij} x_j
\end{equation}
\noindent
so obviously
\begin{equation}
perm(A) = < (\sum_{i_1} A_{1,i_1} \eta_{i_1})
           (\sum_{i_2} A_{2,i_2} \eta_{i_2})...>.
\label{zperm}
\end{equation}

The sum of permanental $k$-minors of a square matrix of size $n$ is
\begin{equation}
    p_k(A) = \sum perm(A_{r, s})
\end{equation}
where $r, s$ are all the order $n-k$ subsets of $(1,..,n)$ and $A_{r, s}$
is the minor obtained eliminating the rows $r$ and the columns $s$.
Using directly this formula\cite{Brualdi},
since the complexity for computing the permanent of $A$ using the Ryser 
algorithm\cite{ryser} is $O(2^n n)$,
the complexity for computing the case $k=n/2$ for $n$ even
is $\binom{n}{n/2}^2 O(2^{n/2} n) \simeq O(2^{\frac{5 n}{2}})$.

The generating function of the sums of permanental minors of the
$m \times n$ matrix $A$ is
\begin{equation}
\sum_k p_k(A) t^k = < (1 + t \sum_{i_1=1}^n A_{1,i_1} \eta_{i_1})
                            (1 + t \sum_{i_2=1}^n A_{2,i_2} \eta_{i_2})...
                            (1 + t \sum_{i_m=1}^n A_{m,i_m} \eta_{i_m})>
\label{permin}
\end{equation}

To compute Eq.(\ref{permin}), after evaluating the $i$-th partial
product, there are $2^n$ monomials in $\eta$, each of them multiplied
by a polynomial of degree at most $i$ in $t$.  Multiplying by $(1 + t
\sum_{j=1}^n A_{i,j} \eta_{j})$ and expanding the product one gets $i
2^n n$ terms, so that the complexity of computing the generating
function for the sum of permanental minors is $O(2^n m^2 n)$.  In all
the estimates of the time complexity, we have neglected the contribution of
number multiplication.

In the case of a square band matrix of size $n$ with entries $M_{ij} = 0$ 
for $|i - j| > w$, at the end of the $i$-th partial product, 
the number of $\eta$-elements is $\nu=min(2w, n)$, so that there are
$2^{min(2w, n)}$ monomials in $\eta$, each multiplied by a polynomial of
degree $i$; 
the computational complexity is $O(2^{min(2w,n)} (w+1) n^2)$ i.e., 
for fixed $w$, it is polynomial in $n$,
while the algorithm in Ref.[\onlinecite{Brualdi}] is exponential.

If one is interested only in computing the permanent using Ryser
algorithm\cite{ryser}, the complexity is $O(n2^n)$, even in the case of
banded matrices, while with our algorithm it is $O(n^2)$.

 If the matrix is ``almost banded'', i.e. it has $h$ non-zero elements
 elements outside the band, with $h$ small, we have to replace $2w$
 with $2w + h$ in the above estimate of the complexity.

 \subsection{Matching generating polynomial}
The  matching generating polynomial of  $G$, i.e.  the generating function
 of the number $N(i)$ of different  matchings of  $G$
containing $i$ edges, is defined by
\begin{equation}
M(t) = \sum_{i=0}^{[v/2]} N(i)t^i
\end{equation}
(Equivalently one defines the matching  polynomial
\begin{equation}
\mu(t) = \sum_{i=0}^{[v/2]}(-1)^i N(i)t^{v - 2i}
\end{equation}
related to the former by $\mu(x) = x^vM(-x^{-2})$)

Consider now Eq.(\ref{prodobj}) in the case of dimers:
\begin{equation}
Z_G = <\prod_{<i,j>}(1 + w_{i,j} \eta_i \eta_j)>
\label{mgp}
\end{equation}
The matching generating polynomial $M(t)$ is obtained from $Z_G$,
setting $w_{i,j} = t$.

\begin{figure}[tbp]
    \begin{center}
        \leavevmode
        \includegraphics[width=5.37 in]{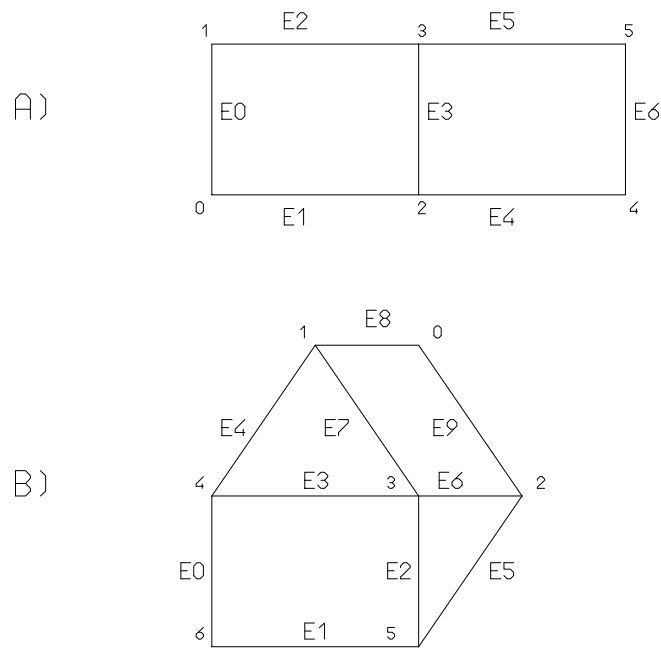}
        \caption{ \label{figura1}  Graphs $A)$ and $B)$.}
    \end{center}
\end{figure}

To make clear by an example the algebraic manipulations used, consider
the graph $A)$ in Fig.\ref{figura1}.

Evaluate the partial product $O_0 O_1$ 
($O_0$ is associated to the edge $E_0$ etc.): one has
\begin{equation}
M_G = <(1 + t \eta_0 \eta_1 + t \eta_0 \eta_2)O_2 O_3 O_4 O_5 O_6> =
<(1 + t \eta_1 + t \eta_2) O_2 O_3 O_4 O_5 O_6>
\nonumber
\end{equation}
In the last step $\eta_0$ has been replaced with $1$ since it
does not occur in the remaining terms $O_2,..,O_6$.  
Similarly in the next step, after expanding the partial product $O_0 O_1 O_2$ 
we can set $\eta_1=1$
\begin{equation}
    M_G = <(1 + t + t\eta_2 + t\eta_3 + t^2\eta_2 \eta_3)O_3 O_4 O_5 O_6>
\nonumber
\end{equation}
In the partial product $O_0 O_1 O_2 O_3$, a term $(t+t^2 )\eta_2\eta_3$
is added.
After expanding the partial product $O_0 O_1 O_2 O_3 O_4$, we can
set $\eta_2=1$.
\begin{equation}
    M_G = <(1 + 2t + (2t+2t^2)\eta_3 + (t+t^2)\eta_4 + t^2\eta_3\eta_4)O_5 O_6>
\nonumber
\end{equation}

In the partial product $O_0 O_1 O_2 O_3 O_4 O_5$, we can set $\eta_3=1$:
\begin{equation}
M_G = <1+4t+2t^2 + (t+2t^2)\eta_4 +(t+2t^2)\eta_5+(t^2+t^3)\eta_4\eta_5)O_6>
\nonumber
\end{equation}
Finally
\begin{equation}
M_G =  1+7t+11t^2+3t^3
\label{es1}
\end{equation}
In each step there are at most two $\eta$ elements.

Given a graph $G$ with $V$ vertices and $E$ edges, 
start with the empty graph $G_0$ on $V$ vertices, 
add an edge to get $G_1$; then continue to add edges, until
$G_E=G$. For a graph $G_i$ in this sequence, an ``active node'' is by
definition a node which is incident with at least one edge, and has a
degree less than the degree that the node has in $G$. The active node
number $\nu$ is the maximum number of active nodes in the sequence
$G_0, ..., G_E$.  
In general the size of the computer memory used by 
  the algorithm grows with a factor $2^{\nu}$.

The graph in the above example has active node number $\nu=2$.

The time complexity for computing the matching polynomial for graphs
with small active node number is $O(2^\nu v^3)$,
analogously to the case of the sums of permanental minors for band matrices;
the space complexity is $O(2^\nu v^2)$.
For fixed $\nu$, the complexity is polynomial in $v$.
Therefore one can deal with large graphs, provided $\nu$ is small.

We have not yet devised a general prescription to determine an
ordering for which $\nu$ is close to minimum.  A simple greedy
procedure to get a sequence with small (but generally non optimal)
$\nu$ is the following: as long as it is possible, add an edge at the
time without increasing the value of $\nu$; otherwise add an edge of
one among the shortest paths in $G - {\it(non-active}$ ${\it
  vertices)}$, which join active vertices.

As an example of a sequence of random graphs with fixed $\nu$,
take a sequence of regular bipartite graphs constructed in the following way.
Let $G_0$ be a cycle with $k$ vertices, with $k$ even. 
Add another cycle with $k$ vertices;
the odd (even) vertices of this cycle are linked respectively
to the even (odd) vertices of the
previous cycle in a random way, obtaining $G_1$;
continue adding cycles in this way, obtaining the sequence ${G_i}$.
The sequence of the corresponding regular bipartite graphs
is obtained linking the vertices of the first and last cycle.
Since the active node number is $\nu=2k$ (the vertices on the first and
the last cycle are active), computing the matching polynomials
for a sequence of $N$ cycle graphs  takes $O(N^2)$.
The code for computing a sequence with $k=6$ is included in the examples
 reported in Ref.[\onlinecite{goh}].

In Ref.[\onlinecite{creutz}] an ordering of fermionic
variables, with the insertion of a projector excluding a fermionic
operator, is similarly chosen. 
Since this formalism is applied to fermions on a finite
square lattice (a grid), there is a natural way to establish a longitudinal
direction and a transverse direction. Only fermionic modes on the transverse
direction appear in the computation, so one can deal with long grids
with few modes in the transverse direction.

Let us observe that
from Eq.(\ref{mgp}), distributing a term
$(1 + w_{k,l} \eta_k \eta_l)$ we get
\begin{equation}
Z_G(t) = <\prod_{<i,j> \neq <k,l>}(1 + w_{i,j} \eta_i \eta_j)> +
w_{k,l} <\prod_{<i,j>, i \ne k, l; j \ne k, l}(1 + w_{i,j} \eta_i \eta_j)>
\label{rec1}
\end{equation}
which gives a recursion relation for matching generating polynomials\cite{godsil}
\begin{equation}
Z_G(t) = M_{G - <k,l>}(t) + t M_{G - k - l}(t)
\end{equation}

If the graph $G$ is bipartite, let us indicate by $y_i$ the elements
$\eta_i$ associated to the even sites, with $\eta_i$ those associated
to the odd sites; then

\begin{equation}
Z_G = < (1 + \sum_{i_1} y_1 w_{1,i_1} \eta_{i_1})
           (1 + \sum_{i_2} y_2 w_{2,i_2} \eta_{i_2})...>
\end{equation}
\noindent
Since the $y_j$ element occurs only in the jth term of the product,
it can be set equal to unity, so that

\begin{equation}
Z_G = < (1 + \sum_{i_1} w_{1,i_1} \eta_{i_1})
           (1 + \sum_{i_2} w_{2,i_2} \eta_{i_2})...>
\label{zbip}
\end{equation}
which gives Eq.(\ref{permin}) in the case $w_{i,j} = t A_{i,j}$.

As an application, we have computed $M(-1)$ for some periodical square
lattices and for some hexagonal lattices in the brick-wall
representation considered in Ref.[\onlinecite{Eerten}]; we disagree
with Ref.[\onlinecite{Eerten}] in some cases, see Appendix A.

\subsection{Independence polynomial}

The independence polynomial $I_G(t) = \sum_i a(i) t^i$ 
is the generating function for the number $a(i)$ of ways of choosing $i$
independent vertices on $G$.
A hard object is made associating to a vertex the
product of the $\eta$ elements on the edges incident with that vertex.
The greedy algorithm for ordering the product consists in choosing a
short path in $G - {\it(non-active}$ ${ vertices)}$.

The matching generating polynomial of a graph $G$ is the independence polynomial
of the {\it line graph} of $G$.

As an example, consider the line graph of the graph $B)$ in Fig.\ref{figura1}.
Evaluate the partial product $O_0 O_1$, set $\eta_8=1$
\begin{equation}
I = I(L(G)) = <O_0...O_6> = <(1 + t\eta_9 + t\eta_4\eta_7)O_2...O_6>
\end{equation}
Evaluate the partial product $O_0 O_1 O_2$, set $\eta_9=1$
\begin{equation}
I = <(1 + t + t\eta_4\eta_7+t\eta_5\eta_6 + t^2\eta_4\eta_5\eta_6\eta_7)O_3...O_6>
\nonumber
\end{equation}
Evaluate the partial product $O_0 O_1 O_2 O_3$, set $\eta_6=\eta_7=1$
\begin{equation}
I = <(1 + t + t\eta_4 + t\eta_5 + (t+t^2)\eta_2\eta_3 + 
(t+t^2)\eta_4\eta_5 )O_4...O_6>
\nonumber
\end{equation}
Evaluate the partial product $O_0 O_1 O_2 O_3 O_4$, set $\eta_3=\eta_4=1$
\begin{equation}
I = <(1+2t+(t^2+t)\eta_0 + (t+t^2)\eta_2 + (t+t^2)\eta_5 + t^2\eta_0\eta_5)O_5 O_6>
\nonumber
\end{equation}
Evaluate the partial product $O_0 O_1 O_2 O_3 O_4 O_5$, set $\eta_2=\eta_5=1$
\begin{equation}
    I = <1+4t+2t^2 + (t+2t^2)\eta_0 + (t+2t^2)\eta_1 + (t^2+t^3)\eta_0\eta_1)O_6>
    \nonumber
\end{equation}
Finally one gets the same as in Eq.(\ref{es1}).

As a check, we computed $I(-1)$ for the hexagonal lattices in the
brick-wall representation considered in table VII of
Ref.[\onlinecite{Eerten}], which can be interpreted as the Witten
index of the quantum hexagonal model.

As another application, we computed $I(1)$ for square grids of size up
to $35 \times 35$.  The results agree with Ref.[\onlinecite{sloane}]
where results are reported up to the size $33 \times 33$.

For the $34 \times 34$ square grid, we get

$I_{34 \times 34}(1) =
387891128933234889019525245048798489818497881776634515543429025520$

$63467216387170202504801083048930878829135642627665925385007961085158
40997971$

$52548773065607505250668587876084152495126750481594564582029827282$

In the $35 \times 35$ case, we get

$I_{35 \times 35}(1) =
72124294712717214286776360359845549941067616972563902046316263757$

$75367684382824803303614885294518590803525311051635720880907913188
1311521646$

$489569003949604822376420723536367577998661984
8116510736835320875797768521522$

$195$

In theoretical chemistry $I(1)$ is called the Merrifield-Simmons
index\cite{MS}.  In Appendix B we have computed the Merrifield-Simmons
index of the Buckminster fullerene.

\section{Conclusions}
We have shown that a simplified version of Creutz algorithm can be
used to compute sums of permanental minors, matching and independence
polynomials. In the case of the sums of permanental minors, we have
shown that this algorithm has lower complexity than using the
Brualdi-Ryser formula.  The algorithms are in general exponential, but
they can become polynomial in particular cases.  For example, sums of
permanental minors have polynomial complexity if the matrix is
banded. It is then important to be able to recognize whether a matrix can
be brought to banded form permuting its rows and its columns.  A
similar ordering problem is met when computing the matching and
independence polynomials. We did not address the problem of finding an
optimal ordering: presumably it is related to the tree decomposition of
graphs\cite{RS}.

\section{Appendix A: The Witten index for rectangular and hexagonal periodic
lattices}

In Ref.[\onlinecite{Eerten}] the Witten index $W=\sum (-1)^i N(i)$ is
evaluated for the supersymmetric dimer model.  For the largest lattices we
agree with these results  only modulo $2^{32}$. We think it likely that in
Ref.[\onlinecite {Eerten}] the large integer arithmetic was
inadequately managed.  
We checked only the cases $m\times n$ for $m, n \ge 4$ and even.
The disagreeing values are listed in Tables
\ref{tab1} and \ref{tab4}.

\begin{table}[ht]
  \caption{Comparison of the values of the Witten index $W(G)$ for a
 square grid $G$ of size $m \times n $ obtained by the algorithm
 introduced in this paper with the disagreeing results in
 table III in Ref.[\onlinecite{Eerten}]. }
\begin{tabular}{|c|c|c|c|c|c|}
 \hline
 $m$ & $n$&$W$ & $|W|^{(1/(m n))}$& $W$ of Ref.[\onlinecite{Eerten}]&
 $|W|^{(1/(m n))}$ in Ref.[\onlinecite{Eerten}] \\
 \hline
10&    8&     -14550253471&    1.340&-1665351583&1.304 \\
10&    10&   3235851927936 &   1.334&1741554048& 1.237 \\
 \hline
 \end{tabular}
 \label{tab1}
\end{table}

We have computed the index $W(G)$ also for the larger lattices indicated in
Table \ref{tab2}.
\begin{table}[ht]
\caption{The value of the Witten index $W$ for a square grid of
size  $m \times n $ larger
than those considered in  Ref.[\onlinecite{Eerten}]. }
\begin{tabular}{|c|c|c|c|}
 \hline
 $m$ & $n$&$W$ & $|W|^{(1/(m n))}$   \\
 \hline
12&    10&    -139080563404700&             1.312\\
12&    12&    988571682202805376&           1.333\\
 \hline
 \end{tabular}
 \label{tab2}
\end{table}
The quantity $|W|^{(1/(mn))}$ should be compared with the expression
 $W = 2r^{m n}cos(m n \theta + \theta_0)$ of
 Ref.[\onlinecite{Eerten}], where $r=1.33 \pm 0.01$

We have compared our evaluations of the Witten indices with the
results in Ref.[\onlinecite{Eerten}] also in the case of hexagonal
lattices. Our results agree only modulo $2^{32}$ with 
Ref.[\onlinecite{Eerten}] (table VIII in that reference);
the disagreeing values are  shown in Table \ref{tab4}.

\begin{table}[ht]
\caption{The values of the Witten index $W$ for  hexagonal grids of
size  $m \times n $, disagreeing
with those considered in  Ref.[\onlinecite{Eerten}]. }
\begin{tabular}{|c|c|c|c|}
 \hline
 $m$ & $n$&$W$& $W$ of Ref.[\onlinecite{Eerten}]  \\
 \hline
10   &  14 &        7711439360&-878495232\\
10   &  16 &        -655517342208&1612654080\\
12   &  12 &        94909515776&420235264\\
12   &  14 &        6459966411264&335598080\\
12   &  16 &        100182729294336&-1677852160\\
14   &  10 &        11948085184&-936816704\\
14 &    12 &        6736033699456&1524979328\\
14  &   14  &       742553681809408&1080971264\\
14   &  16 &        -1384901745575424&1869085184\\
 \hline
 \end{tabular}
 \label{tab4}
\end{table}

We have computed the index $W$ also for the larger lattice indicated in
Table \ref{tab3}.
\begin{table}[ht]
\caption{The value of the Witten index $W$ for hexagonal grids of
size  $m \times n $ larger
than those considered in  Ref.[\onlinecite{Eerten}]. }
\begin{tabular}{|c|c|c|}
 \hline
 $m$ & $n$&$W$   \\
 \hline
16& 14   &-119633551609600            \\
16& 16   &-6060918132969537536\\
 \hline
 \end{tabular}
 \label{tab3}
\end{table}
For $10 \le m, n \le 16$ the Witten index per site $|W|^{(1/(mn))}$
is between $1.156$ and $1.192$ with an average value $1.18$, which
lies below the interval $r=1.4 \pm 0.1$
 reported in Ref.[\onlinecite{Eerten}].

\section{Appendix B: Usage of the Python module ``hobj''}
The module ``hobj'' can be downloaded from Ref.[\onlinecite{goh}].
It can be used without any dependence; the code for univariate polynomials,
represented as arrays, is adapted from SymPy{\cite{sympy}.
Here is the example  graph $A)$ in Figure \ref{figura1} 

\begin{verbatim}
>>> from hobj import dup_matching_generating_poly
>>> d = {0:[1,2], 1:[0,3], 2:[0,3,4], 3:[1,2,5], 4:[2,5], 5:[3,4]}
>>> dup_matching_generating_poly(d)
[3, 11, 7, 1]
\end{verbatim}

Since it is a bipartite graph, one can compute it also using the
reduced adjacency matrix
\begin{verbatim}
>>> from hobj import dup_permanental_minor_poly
>>> from domains import ZZ
>>> m = [[1,1,0],[1,1,1],[0,1,1]]
>>> dup_permanental_minor_poly(m, ZZ)
[3, 11, 7, 1]
\end{verbatim}
In the case of bipartite graphs the second way is often faster.

The following examples take a fraction of a second on a current
personal computer.
Compute the sum of the permanental minors of a banded matrix
\begin{verbatim}
>>> from hobj import dup_permanental_minor_poly
>>> from domains import ZZ
>>> m = [[i*j if abs(i-j) < 6 else 0 for i in range(20)] for j in range(20)]
>>> sum(dup_permanental_minor_poly(m, ZZ))
11936810897247956264161397956481650508142206788L
>>> dup_permanental_minor_poly(m, ZZ, 1)
11936810897247956264161397956481650508142206788L
\end{verbatim}

the second way is faster and uses less memory because it avoids 
constructing the polynomial. Similarly in the following examples.

Let us call ``hobj'' from Sage\cite{Sage} and compute the sum of the
coefficients of the matching polynomial for the Buckminster fullerene $C_{60}$
(truncated icosahedron)
computed first in Ref.[\onlinecite{hosoya}]

\begin{verbatim}
sage: from hobj import dup_matching_generating_poly
sage: d = graphs.BuckyBall().to_dictionary()
sage: sum(dup_matching_generating_poly(d))
1417036634543488
sage: dup_matching_generating_poly(d, val=1)
1417036634543488
\end{verbatim}

Same for the independence polynomial
\begin{verbatim}
sage: from hobj import dup_independence_poly
sage: sum(dup_independence_poly(d))
217727997152
sage: dup_independence_poly(d, val=1)
217727997152
\end{verbatim}

\end{document}